\documentclass[sigconf,screen,nonacm]{acmart}

\graphicspath{ {./img/} }
\usepackage{microtype}
\usepackage{balance}
\usepackage{enumitem}
\usepackage{makecell}
\usepackage{subcaption}
\usepackage{fontawesome5}
\usepackage{framed}
\usepackage{xcolor}
\usepackage{siunitx}
\colorlet{promptshade}{gray!10}
\colorlet{promptborder}{black!30}
\newenvironment{promptbox}{%
  
  \MakeFramed {
      \advance
      \hsize-\width
      \FrameRestore}%
      \setlength{\FrameSep}{4pt} 
      \raggedright 
      \ttfamily
}{
  \endMakeFramed
}

\colorlet{testimonialshade}{yellow!10}
\colorlet{testimonialborder}{black!30}
\newenvironment{testimonialbox}{%
  \setlength{\FrameSep}{3pt} 
  \MakeFramed {%
    \advance\hsize-\width
    \FrameRestore}%
  \setlength{\parindent}{2pt} 
  \raggedright 
  \ttfamily \small
}{%
  \endMakeFramed
}

\usepackage[nolist]{acronym} 
\makeatletter
\let\orig@ac\ac
\renewcommand{\ac}[1]{%
  \@ifundefined{AC@#1}{%
    \PackageError{acronym}{Undefined acronym `#1`}{%
      Define it using \string\acro{#1}{...}%
    }%
  }{%
    \orig@ac{#1}%
  }%
}
\makeatother

\begin{acronym}

\newacro{GoogleUSA}[Google]{}
\newacro{AutoRepair}[\texttt{FlowAgent}]{}
\newacro{AutoRepairLaunchDate}[October 2025]{}
\newacro{TotalEvalDevelopers}[\num{3}]{}
\newacro{TotalEvalTeams}[\num{37}]{}
\newacro{TotalEvalExamples}[\num{195}]{}
\newacro{TotalEvalExamplesGood}[\num{131}]{}
\newacro{TotalEvalExamplesBad}[\num{64}]{}
\newacro{TotalEvalAccuracy}[$67.18\%$]{}
\newacro{TotalInterviewedDevelopers}[\num{9}]{}
\newacro{DeveloperA}[\texttt{P-1}]{}
\newacro{DeveloperB}[\texttt{P-2}]{}
\newacro{DeveloperC}[\texttt{P-3}]{}
\newacro{DeveloperD}[\texttt{P-4}]{}
\newacro{DeveloperE}[\texttt{P-5}]{}
\newacro{DeveloperF}[\texttt{P-6}]{}
\newacro{DeveloperG}[\texttt{P-7}]{}
\newacro{DeveloperH}[\texttt{P-8}]{}
\newacro{DeveloperI}[\texttt{P-9}]{}

\newacro{FixAttemptLatencyP10}[\num{2.37} minutes]{}
\newacro{FixAttemptLatencyP25}[\num{6.72} minutes]{}
\newacro{FixAttemptLatencyP50}[\num{27.48} minutes]{}
\newacro{AutoRepairLatencyP10}[\num{4.78} minutes]{}
\newacro{AutoRepairLatencyP50}[\num{9.85} minutes]{}
\newacro{AutoRepairLatencyP90}[\num{25.83} minutes]{}
\newacro{AutoRepairLatencyP95}[\num{30} minutes]{}

\newacro{sym:funnel}[$\searrow$]{}
\newacro{TotalChanges}[\num{2567829}]{}
\newacro{TotalChangesDroppedPreExecution}[\num{1785955}]{}
\newacro{TotalChangesAttempted}[\num{781874}]{}
\newacro{TotalChangesAttemptedPct}[$30.45\%$]{}
\newacro{TotalChangesAttemptedSuccessful}[\num{421818}]{}
\newacro{TotalChangesAttemptedSuccessfulPct}[$53.95\%$]{}
\newacro{TotalChangesDroppedPostExecution}[\num{126310}]{}
\newacro{TotalChangesPostedTo}[\num{295508}]{}
\newacro{TotalChangesPostedToPct}[$70.06\%$]{}
\newacro{TotalChangesPreviewed}[\num{65069}]{}
\newacro{TotalChangesPreviewedPct}[$22.02\%$]{}
\newacro{TotalChangesApplied}[\num{28554}]{}
\newacro{TotalChangesAppliedPct}[$43.88\%$]{}
\newacro{TotalChangesPreviewedCritique}[\num{55029}]{}
\newacro{TotalChangesAppliedCritique}[\num{20826}]{}
\newacro{CritiqueAppliedOverPreviewed}[$37.84\%$]{}
\newacro{TotalChangesPreviewedCider}[\num{21279}]{}
\newacro{TotalChangesAppliedCider}[\num{12702}]{}
\newacro{CiderAppliedOverPreviewed}[$59.69\%$]{}
\newacro{TotalAuthors}[\num{36479}]{}
\newacro{ChangesMeanFiles}[\num{10.57}]{}
\newacro{ChangesMedianFiles}[\num{6}]{}
\newacro{ChangesTotalFileExtensions}[\num{917}]{}
\newacro{ChangesMeanTests}[\num{16.49}]{}
\newacro{ChangesMedianTests}[\num{2}]{}
\newacro{ChangesMeanToolCalls}[\num{21.37}]{}
\newacro{ChangesMeanTokens}[\num{424149}]{}
\newacro{ChangesTotalTokens}[\num{643} billion]{}

\end{acronym}

\begin{document}
\title{Catching Developers in the Flow: Low-Latency Agentic Program Repair at Google Scale}

\author{Celal Ziftci}
\correspondingauthor
\orcid{0009-0003-2951-8698}
\affiliation{%
  \institution{Google}
  \city{New York}
  \country{USA}
}
\email{ziftci@google.com}

\author{Spencer Greene}
\orcid{0009-0007-0240-5377}
\affiliation{%
  \institution{Google}
  \city{New York}
  \country{USA}
}
\email{greenespencer@google.com}

\author{Ray Liu}
\orcid{0009-0009-7183-4743}
\affiliation{%
  \institution{Google}
  \city{New York}
  \country{USA}
}
\email{rayliu@google.com}

\author{Livio Dalloro}
\orcid{0009-0008-7565-6406}
\affiliation{%
  \institution{Google}
  \city{New York}
  \country{USA}
}
\email{dalloro@google.com}

\author{Lorenzo Dini}
\orcid{0009-0004-6334-6735}
\affiliation{%
  \institution{Google}
  \city{New York}
  \country{USA}
}
\email{dini@google.com}

\begin{abstract}

Manual repair of program failures is time-consuming and disruptive for software developers, particularly during the pre-submit phase where test failures occur within continuous integration systems. While Automated Program Repair has seen significant advancement through Large Language Models, existing state-of-the-art techniques primarily focus on post-submit workflows, operating offline without the low-latency requirements necessary to assist developers in real-time within their flow before they switch context.

In this paper, we introduce \acs{AutoRepair}, an AI agent deployed at Google to automatically repair test failures in the pre-submit outer-loop workflow inside continuous integration systems. Integrated into Google’s internal developer tools, Critique and Cider, \acs{AutoRepair} utilizes a ReAct-style generate-and-validate loop, as well as rigorous pre-execution and post-execution abstention filters to ensure high-quality suggestions under strict latency constraints.

Based on our case studies, \acs{AutoRepair} is highly effective. First, a manual evaluation conducted on \acs{TotalEvalExamples} real-world test failures demonstrated \acs{TotalEvalAccuracy} accuracy in suggesting correct fixes. Following its Google-wide deployment, \acs{AutoRepair} suggested fixes on \acs{TotalChangesPostedTo} changes, of which developers previewed \acs{TotalChangesPreviewed} and applied \acs{TotalChangesApplied}. Developer feedback from interviews indicate that the agent is useful in suggesting correct fixes, integration of autonomous repair agents into industrial software engineering workflows is received well, while interesting challenges and opportunities still remain.
\end{abstract}



\maketitle

\section*{Preprint Notice}
This is a preprint of a paper accepted at the 41st IEEE/ACM International Conference on Automated Software Engineering (ASE 2026).

\section{Introduction}

Software developers frequently modify codebases to add new features, fix bugs, or improve system security and performance. To ensure these modifications do not introduce regressions into the wider codebase, they must be rigorously tested. When code modifications result in test failures, developers must engage in program repair. This process is time-consuming and disruptive, particularly in the outer-loop pre-submit phase, where test failures occur outside the developers' integrated development environments (IDE) within the continuous integration systems, like Google's TAP \citep{tap1,tap2}, to identify global regressions before a change is submitted.

Upon notification of a failure, developers typically investigate the failure and implement fixes. Our analysis reveals that the median time between a failure notification and a developer's subsequent code edit at Google is \acs{FixAttemptLatencyP50}. This provides a critical window where automated repair assistance could meaningfully improve developer velocity and reduce frustration.

To alleviate the burden of manual debugging, the field of Automated Program Repair (APR) has received significant attention. Early APR solutions relied on manually described mutation rules, symbolic constraints, and static analysis. More recently, the advent of Large Language Models (LLMs) has driven a shift toward machine learning and autonomous agent-based APR. State-of-the-art tools such as SWE-Agent \citep{yang2024swe}, AutoCodeRover \citep{zhang2024autocoderover}, Google's Passerine \citep{rondon2025evaluating} and Meta's Engineering Agent \citep{maddila2025agentic} utilize ReAct-style \citep{yao2023react} loops to autonomously navigate codebases, utilize tools, and generate fixes. However, these approaches primarily focus on the post-submit workflow, designed to repair programs in an offline capacity, operating without latency constraints required to assist developers in real-time before they switch context out of their flow.

In this paper, we introduce \acs{AutoRepair}, an AI agent deployed at Google to automatically debug and repair test failures directly within the pre-submit outer-loop workflow. Successful fixes, alongside an AI-generated summary of the root cause and the fix, are surfaced in Google's internal development tools, allowing developers to preview and apply fixes inside their habitual workflows.

To the best of our knowledge, unlike post-submit APR tools, \acs{AutoRepair} is the first APR system in the pre-submit outer-loop workflow built to operate under critical latency requirements to catch developers in their active change modification flow.

We provide a comprehensive empirical evaluation of \acs{AutoRepair} with, to the best of our knowledge, the largest industrial dataset reported to date, starting first with a manual evaluation, then continuing over a multi-month period in production. We discuss \acs{AutoRepair}'s success in obtaining valid fixes for thousands of failures, and present detailed telemetry and qualitative insights derived from adoption and user interviews, highlighting developer preferences, tooling adoption behaviors, and the varying levels of trust developers place in AI-generated code repairs.

\section{Manual Program Repair}
Developers modify code for many reasons, e.g. to fix bugs, to add new features, to improve performance or security. Once code is modified, in one or more files, developers create a \textit{change}, a mutation to the code repository, reviewed and submitted atomically.

\subsection{Program Failure Stages}

Program failures may involve \textit{build failures} where the code does not compile, or \textit{test failures} where the program builds correctly but there are failing tests. Our work focuses on test failures only.

In a change, when the modified code has accompanying tests that fail, either the program or the tests themselves may need to be updated. Such repairs may happen in two separate development stages:

\begin{enumerate}[leftmargin=*, label=(\arabic*)]
    \item \textbf{Pre-submit}: These repairs happen before the change is submitted, typically as part of the change itself, without impacting other developers.
    \item \textbf{Post-submit}: These repairs happen after breaking changes have been submitted to the repository, in turn impacting all developers intending to modify the code, as there are now failing tests regardless of any new changes.
\end{enumerate}

\noindent Our work focuses on the pre-submit workflow, before faulty code or tests are submitted to the repository.

During pre-submit, automated testing typically follows two distinct workflows depending on the stage of the change life-cycle:

\begin{enumerate}[leftmargin=*, label=(\arabic*)]
    \item  \textbf{Inner-loop}: This is the code modification process within the integrated development environment (IDE). In this phase, either the developer or an AI coding agent, like Antigravity \cite{antigravity}, runs unit tests related to the modified code, providing early feedback before a change is even created. The tests executed in this workflow are typically either adjacent or in close proximity to the edited code, e.g. in the same or the parent directory.
    \item \textbf{Outer-loop}: This is the stage where the local changes are complete and a change has been created. This phase focuses on global correctness, finding and executing all dependent tests in the entire monorepo to ensure that the change does not introduce regressions into the wider codebase across Google.
\end{enumerate}

\noindent Our work focuses on the outer loop, where various tools at Google are used by developers during their program repair workflows, discussed below.

\subsection{Outer-Loop Pre-Submit Developer Tools}
\label{sec:prior-sota}
There are several developer tools used at Google during the outer-loop pre-submit workflow to prepare a change for submission.
\\

\noindent \textbf{Cider}: This is Google's internal IDE \cite{cider}. Developers typically start code changes in Cider, utilize more tooling to help iterate on code changes, and move between Cider and other tooling as they finalize their changes.
\\

\noindent \textbf{Critique}: This is the internal web-based code review system at Google \cite{sadowski2018modern}. After a developer changes code in Cider, they create a change with the modifications. Before sending the change to other developers for review, automated tools and regression tests run on the changed version of the code to help the owner with \textit{findings} and \textit{suggested fixes} by showing notifications on relevant parts of the code, or the entire change itself.
\\

\begin{figure}
\centering
\includegraphics[scale=0.08]{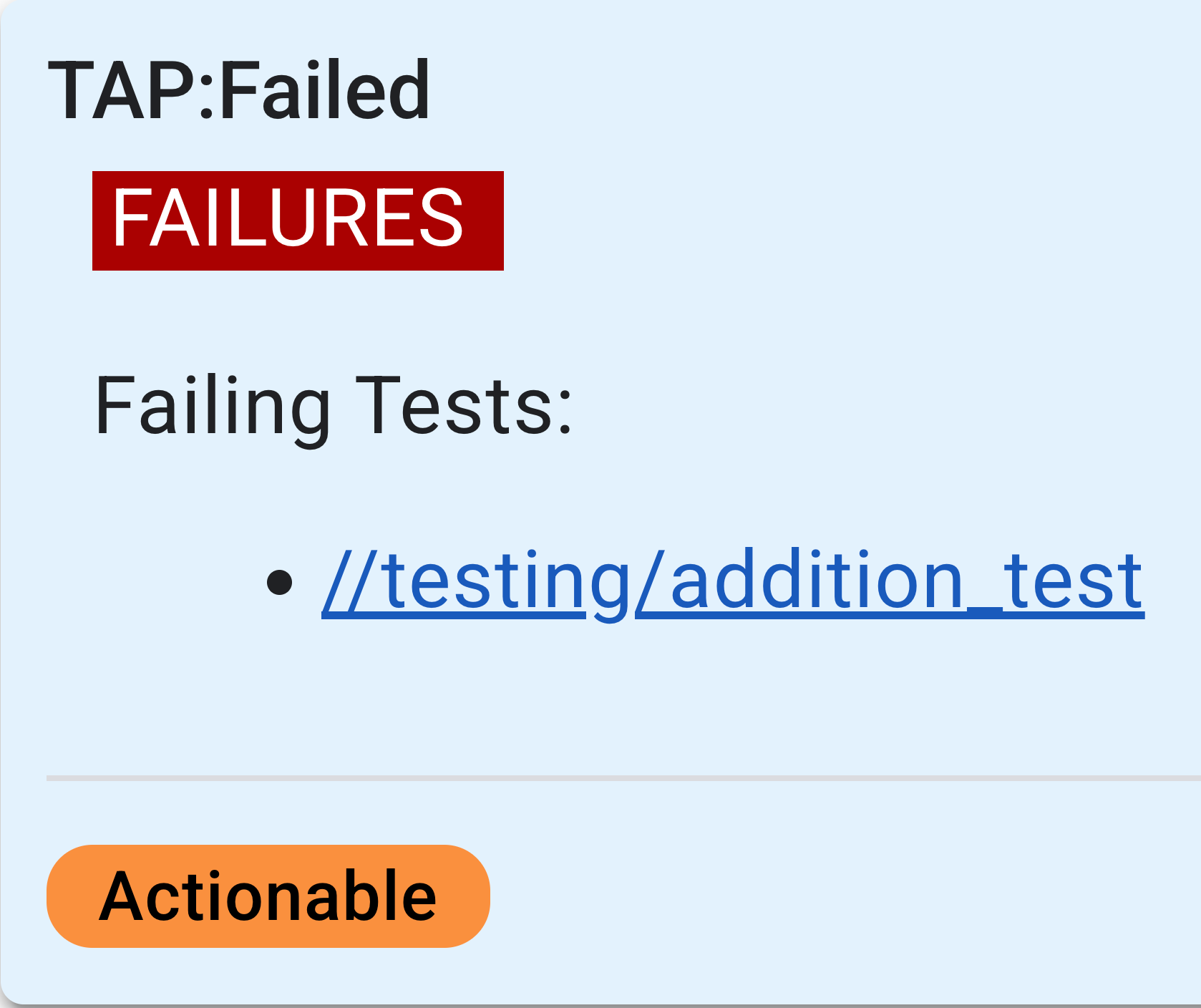}
\Description{Critique finding for failed tests}
\caption{Test failure \textit{finding} surfaced in Critique \citep{sadowski2018modern}, Google's web-based internal code review system.}
\label{fig:critique-test-failed}
\end{figure}

\noindent \textbf{TAP}:  The Testing Automation Platform (TAP) is the continuous integration system at Google  that runs unit-tests \cite{tap1,tap2}. For a change, TAP finds all tests that depend on the change in the monorepo and runs them to identify potential regressions before the change can be submitted. An example TAP test failure finding is shown in Critique in Figure \ref{fig:critique-test-failed}.
\\

Developers typically start coding in Cider, as they get ready to finish their modifications, they create a change and run analyses in Critique, at which time TAP may uncover failing tests due to the changes. Once notified of failures, developers then investigate and attempt to fix the failing tests by modifying their change or the tests themselves in Cider, manually or using assistive tooling.

\begin{figure}
\centering
\includegraphics[scale=0.37]{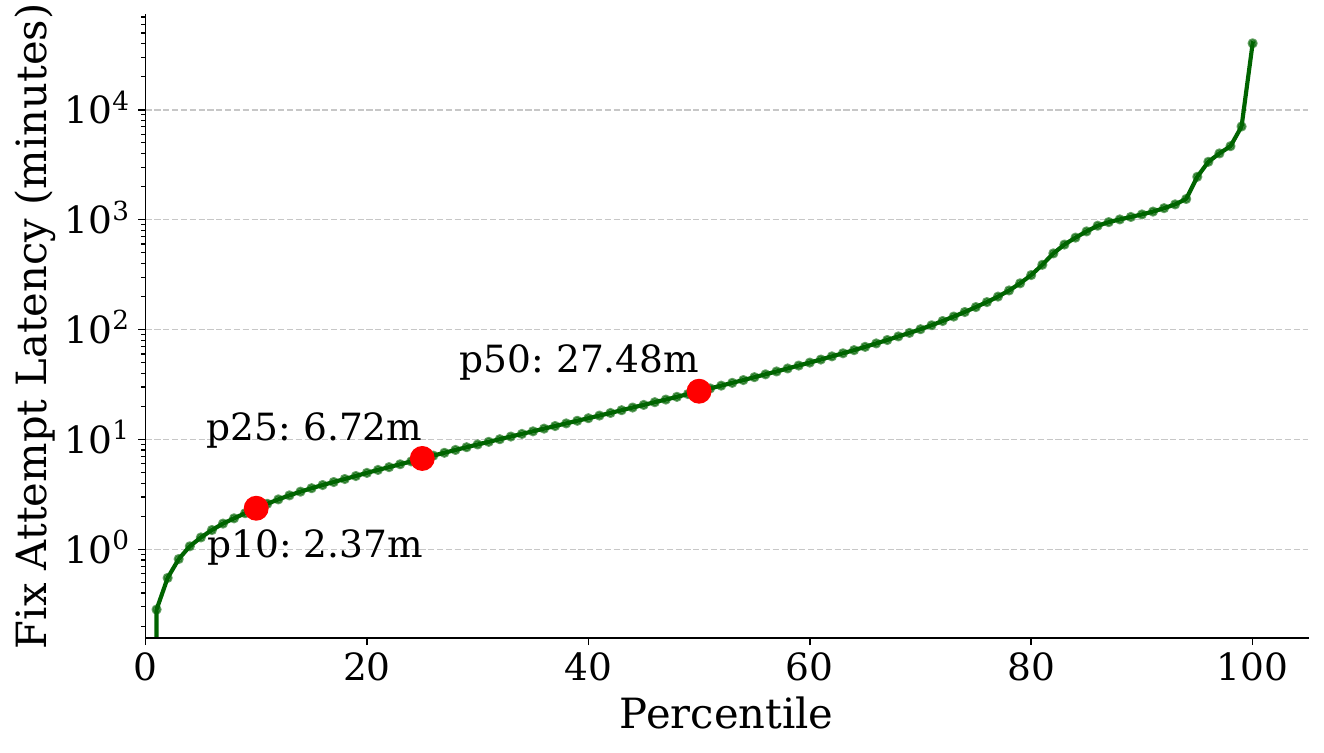}
\Description{Cider suggested fix for failed tests}
\caption{Fix attempt latency distribution.}
\label{fig:fix-attempt-latency}
\end{figure}

We conducted an analysis of code changes submitted across Google within a 30-day period during the month of September-2025 to understand developer behavior on failing tests. We investigated \textit{fix attempt latency}, the wall clock time between TAP reporting test failures on a version of the change and a developer making any additional edits and moving the change to the following version, summarized in Figure \ref{fig:fix-attempt-latency}.

Fix attempt latency is the opportunity window to help developers with test failures, either by providing root cause analysis information or suggesting a potential fix, before they attempt to keep working on the change to fix the tests themselves, and it is \acs{FixAttemptLatencyP25} at p25, and \acs{FixAttemptLatencyP50} at p50.

\section{Automated Program Repair With \acs{AutoRepair}}
\label{sec:solution}
With the recent advances in AI, autonomous AI coding agents have shown great success on automated program repair (APR). To help developers in program repair, we propose \acs{AutoRepair}, an autonomous agent integrated into Google's developer tooling to automatically debug and repair pre-submit outer-loop test failures.

Although \acs{AutoRepair} and the tools it is integrated with are internal to Google, the logic and flow of \acs{AutoRepair} can be replicated using standard developer tooling available in the industry.

\begin{figure*}
\centering
\includegraphics[scale=0.92]{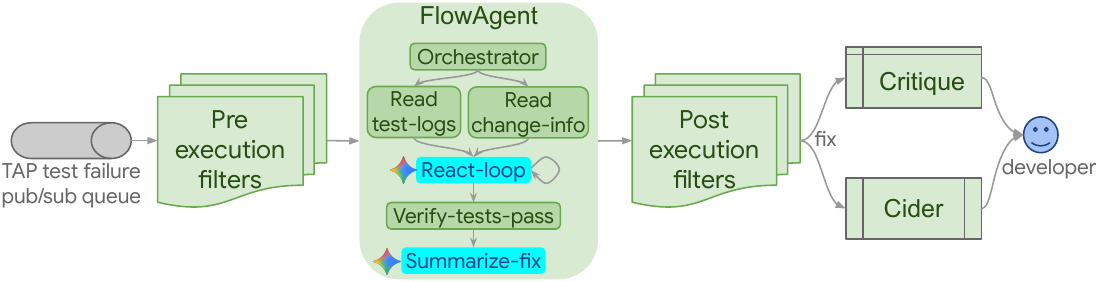}
\Description{\AutoDebug{} design}
\caption{System overview of automatically repairing programs with \acs{AutoRepair}.}
\label{fig:auto-repair-design}
\end{figure*}

The high level architecture of \acs{AutoRepair} is shown in Figure \ref{fig:auto-repair-design}. After a change is created and test execution is triggered in Critique, \acs{AutoRepair} listens to TAP notifications on test failures. Upon failure, without developer involvement, it analyzes the change, checks if it passes supported filters, runs an AI-agent to fix the code, checks if the fix passes safety filters, and finally publishes its fix to Critique for developers to preview and apply either in Critique or Cider.

\subsection{Pre-Execution Filters}
\acs{AutoRepair} ignores certain changes with failure by using pre-execution filters to avoid nonviable or unnecessary fix attempts, and developer confusion due to unrelated or low quality fix suggestions.

These filters, listed below, check characteristics of the change or the failing tests, and are created based on institutional knowledge, manual evaluations and observations on a set of failures, and developer feedback in production.
\\

\noindent
\begingroup
\setlength{\tabcolsep}{2pt}
\begin{tabular}{@{} p{0.27\columnwidth} | p{0.69\columnwidth} @{}}
\makecell[r]{Build failure} & Failure is due to build, not tests.\\
\makecell[r]{Already failing} & Tests already failing in the repository are skipped.\\
\makecell[r]{Flaky} & Failing known flaky tests \citep{fowler,MiccoFlakyGoogle,ziftci2020flake}, those that pass and fail at the same code version without changes, are skipped.\\
\makecell[r]{Automated} & Failing tests on changes created by automated tools are skipped.\\
\makecell[r]{Stale} & Failing tests on changes that the developer already modified are skipped.\\
\makecell[r]{Unsupported files} & Failing tests that involve certain file types, e.g. images and binary files, are skipped.\\
\makecell[r]{Too many files} & Failing tests on changes with more than $100$ files are skipped.\\
\makecell[r]{Too many lines} & Failing tests on changes with more than $1000$ lines are skipped
\end{tabular}
\endgroup
\\

If a test failure on a change passes all filters, \acs{AutoRepair} is triggered.

\subsection{Agentic Repair}
Once a test failure passes the pre-execution filters, the \acs{AutoRepair} hybrid system with predefined and fully agentic steps running in a sandbox, shown in Figure \ref{fig:auto-repair-design}, is triggered with the setup parameters:

\begin{itemize}
    \item LLM = Internal version of Gemini 2.5 Pro \cite{comanici2025gemini} fine-tuned on internal Google code and data \cite{didact,frommgen2024resolving},
    \item $temperature = 0.1$ so that agent actions are mostly deterministic for easy debugging and reproducibility,
    \item $top_p = 0.95$ so that there is room for creativity while keeping out extremely unlikely outcomes.
\end{itemize}

First, the \textit{orchestrator} initiates the agent by reading test logs and reading change information, including its description, the list of modified files and the modifications on them in diff \cite{diff, hunt1976algorithm} format.

These pieces of information are then passed to a ReAct-style \cite{yao2023react} \textit{generate-and-validate} loop using the prompt shown in Listing \ref{fig:gemini-prompt}, where it can iteratively refine its fix based on build and test feedback, with:

\begin{itemize}
    \item Access to tools, including reading/editing/deleting files, searching code, building code, reading build status, reading build logs, executing tests, reading test status and reading test logs,
    \item A maximum of 100 tool calls, to avoid excessive runtime and tool calling,
    \item A maximum runtime of 30 minutes, in line with fix attempt latency discussed in Section \ref{sec:prior-sota} to avoid running for too long and to catch developers in their coding flow.
\end{itemize}

If the loop finishes successfully and produces a fix diff, we conduct a final validation step, either by confirming that the agent executed the tests in its last validation cycle, or by executing the tests and checking the result again. This additional validation is necessary, as LLMs are known to hallucinate, and without final deterministic validation, there is a risk of showing a potentially invalid fix to developers.

If the verification is successful, we summarize the failure root cause and the fix by making a single call to the LLM and passing the transcript of the entire trajectory of the generate-and-validate loop above, with the prompt shown in Listing \ref{fig:gemini-summarization-prompt}.

Finally, the suggested fix and its summary is passed to post-execution filters.

\begin{figure*}[tbp]
\centering
\renewcommand{\figurename}{Listing}

\begin{minipage}{\linewidth}
\centering
\begin{promptbox}
The following change has failing pre-submit tests.\\[0.75em]

The change information is provided below:\\[0.75em]

=== Change description ===\\
\{change\_description\}\\[0.75em]

=== Change diff ===\\
\{change\_diff\}\\[0.75em]

Your task is to analyze the provided error logs from failing tests (paying close attention to +/- diffs indicating discrepancies), and the content of the relevant test file(s).\\[0.75em]

Based on this analysis, you must output only a list of specific modifications required to fix the failing tests, along with their target file names.\\[0.75em]

Your Analysis Steps (Internal Process):\\
- Identify Test Failures \& Discrepancies: Parse the error logs to pinpoint failing tests, error messages, stack traces. Crucially, analyze any + (expected) / - (actual) diffs within the logs, as these often directly indicate necessary corrections in test expectations or data.\\
- Contextualize with Code: Analyze the test files (present in the error logs) to understand the structure, setup, assertions, and logic of the failing test(s) and related code.\\
- Determine Required Modifications: Synthesize all information to identify the exact changes needed. This might involve updating expected values in tests based on log diffs. Do not add any comments or whitespaces.\\[0.75em]

Your Output:\\
- Your output must consist exclusively of the modification instructions and target files in diff format.\\
- Do not include any explanations, analysis summary, greetings, apologies, or conversational text.\\[0.75em]

Fix the following tests that produced errors:\\[0.75em]

=== Test name: \{test\_name\} ===\\[0.75em]
\{test\_error\}
\end{promptbox}

\Description{Agent ReAct loop prompt}
\subcaption{The system prompt template used in the ReAct agent loop.}
\label{fig:gemini-prompt}
\end{minipage}
    
\vspace{0.3cm}

\begin{minipage}{\linewidth}
\centering
\begin{promptbox}
You are an expert software engineer. You are given a transcript of an AI
agent's thoughts and actions to repair a test failure. Please provide a
concise summary of what caused the test failure and the steps taken by the
agent to repair the failure.\\[0.75em]

Try your best to summarize the steps taken by the agent to fix the test.\\
Focus on the specific actions taken and the thought process behind them.\\
Do not simply copy the transcript of repair steps.\\
When possible it's best to include artifacts like file names, method headers, variables, etc.\\[0.75em]

The resulting synopsis should be at most 5 sentences and suitable for a
user interface, but ideally it can be one or two.\\
If appropriate, presenting the synopsis in bullet points can be helpful.\\[0.75em]

It is imperative that the summary follows HTML formatting, so make sure
to use HTML as your only formatting style. Do not use any other formatting.
The summary should be as compact as possible, this means the HTML should be
written in a single line.\\[0.75em]

=== TRANSCRIPT ===\\
\{agent\_transcript\}
\end{promptbox}
\Description{Agent summarization prompt}
\subcaption{The prompt template to summarize the failure root cause and the suggested fix.}
\label{fig:gemini-summarization-prompt}
\end{minipage}

\caption{Prompts used by \acs{AutoRepair}.}
\end{figure*}

\subsection{Post-Execution Filters}

\acs{AutoRepair} uses post-execution filters, listed below, to avoid posting fix suggestions to developers if the fix will not be useful.
\\

\noindent
\begingroup
\setlength{\tabcolsep}{2pt}
\begin{tabular}{@{} p{0.16\columnwidth} | p{0.81\columnwidth} @{}}
\makecell[r]{Stale} & If the developer modified the change while the agent was processing the failing version, the agent's fix is obsolete.\\
\makecell[r]{Submitted} & If the developer already submitted the change, the agent's fix is obsolete.
\end{tabular}
\endgroup
\\

Once the fix passes these filters, it is ready to be displayed to developers.

\subsection{Showing Fixes to Developers}
Developers have varying preferences on the environment where they tend to preview and apply fixes. Some developers prefer using Critique, as the fixes are posted there first, while others prefer going back to the Cider IDE, as they are more used to making changes there. To maximize developer adoption, \acs{AutoRepair} is integrated into both Critique and Cider.

\acs{AutoRepair} fixes are displayed to developers in Critique as part of the TAP test failure notification, shown in Figure \ref{fig:critique-test-failed}, instead of an additional finding, to help developers associate the fix to the failure, and to avoid overwhelming them with too many findings posted on their change, as there are hundreds of analyses executed for each change and several findings may be displayed.

\begin{figure}
\centering
\includegraphics[scale=0.06]{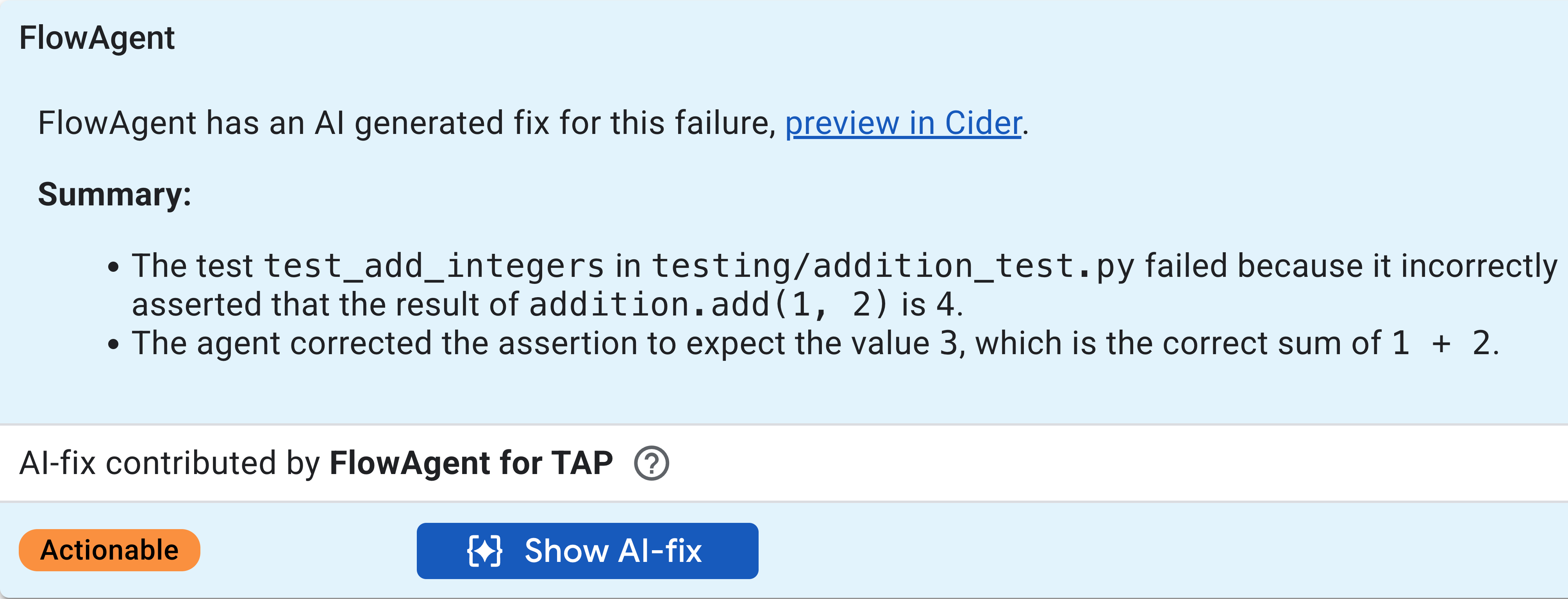}
\Description{\acs{AutoRepair} finding for failed tests}
\caption{\acs{AutoRepair} finding surfaced in Critique, with a root cause summary, and a link to Cider.}
\label{fig:critique-agent-finding}
\end{figure}

Shown in Figure \ref{fig:critique-agent-finding}, the fix contains a summary of the root cause of the failure for developers to first understand if the diagnosis is correct, before viewing and deciding to apply the suggested fix. Additionally, we provide a link to preview the fix in the Cider IDE, and a button to preview and apply it inside Critique.

\begin{figure}
\centering
\includegraphics[scale=0.06]{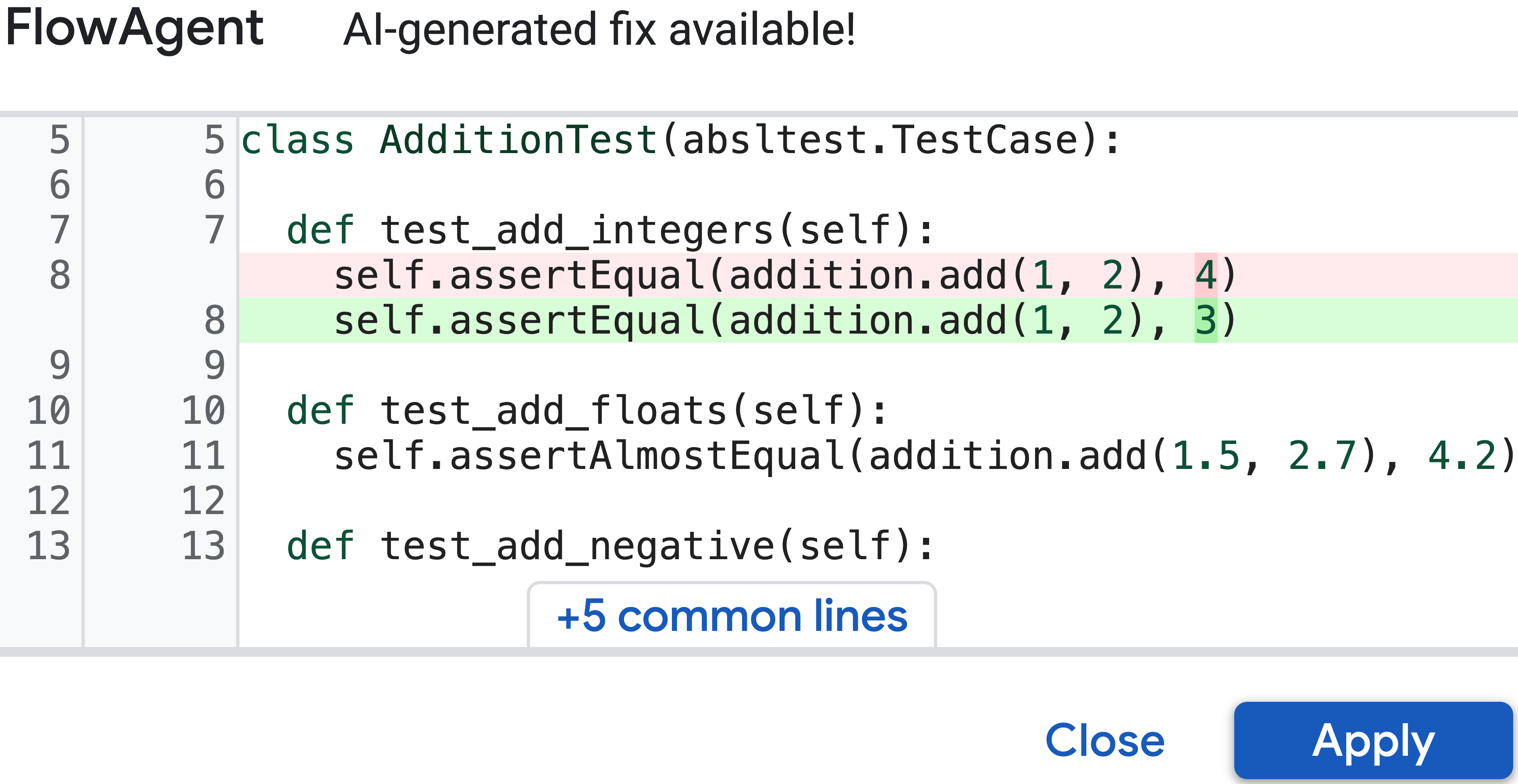}
\Description{Cider suggested fix for failed tests}
\caption{\acs{AutoRepair} suggested fix displayed in Critique.}
\label{fig:critique-agent-suggested-fix}
\end{figure}

Developers can click the button to preview the fix in Critique, as shown in Figure \ref{fig:critique-agent-suggested-fix}, and apply it in place.

\begin{figure}
\centering
\includegraphics[scale=0.270]{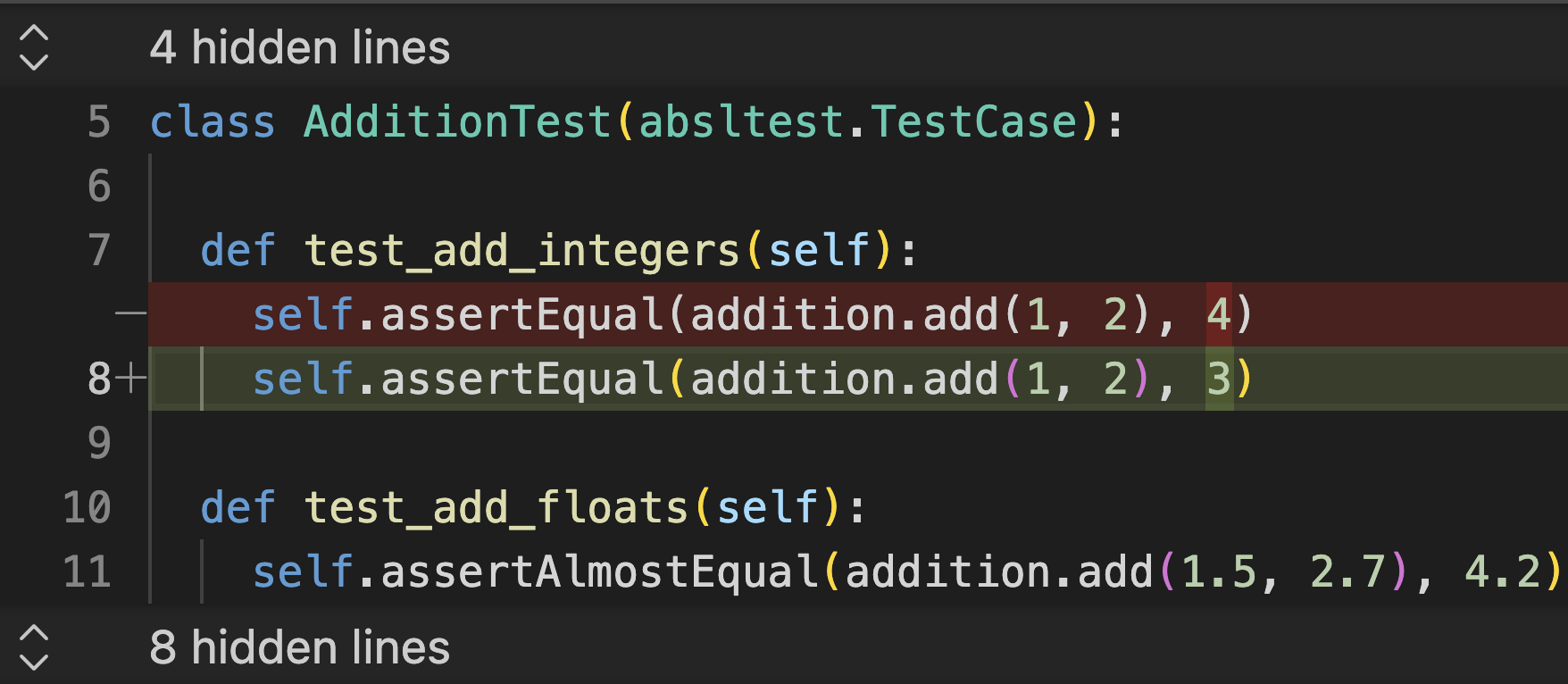}
\Description{Cider suggested fix for failed tests}
\caption{\acs{AutoRepair} suggested fix displayed in Cider.}
\label{fig:cider-agent-suggested-fix}
\end{figure}

For more complex fixes that span several lines or files, or if they want to perform further manual edits, developers may choose to preview and apply the fix in the Cider IDE, as shown in Figure \ref{fig:cider-agent-suggested-fix}.

\section{Evaluation and Discussion}
\label{section:evaluation}
In this section, we discuss how we evaluate our system, describe our case studies, and report the results of our interviews on the usability of our system.

\subsection{Manual Evaluation}
To evaluate the effectiveness of \acs{AutoRepair}, we first conducted a case study where we ran \acs{AutoRepair} on \acs{TotalEvalExamples} randomly selected TAP test failures from \acs{TotalEvalTeams} distinct teams across Google, summarized in Table \ref{table:manual-eval-stats}. We asked \acs{TotalEvalDevelopers} expert developers with at least five years of software development experience each to assess the suggested fixes, similar to shown in Figure \ref{fig:critique-agent-suggested-fix}, and report whether the fix was appropriate for the intent of the change and the failure. We then held a meeting to go over their reports and align on any disagreements, to obtain final agreement on the assessment for each fix.

Based on the experts’ reporting, \acs{AutoRepair} was successful in suggesting correct fixes in \acs{TotalEvalExamplesGood} of the failures, resulting in a \acs{TotalEvalAccuracy} success rate. For the remaining 64 cases, the fixes were either incorrect or not exactly correct, discussed below with examples.

\begin{table}
\centering
\caption{Statistics about the manual evaluations.}
\begin{tabular}{r|l}
\# teams that owned the TAP failures evaluated & \acs{TotalEvalTeams} \\
\# developers that conducted evaluations & \acs{TotalEvalDevelopers} \\
\# failures evaluated & \acs{TotalEvalExamples} \\
\# accurate fixes on the failures evaluated & \acs{TotalEvalExamplesGood} (\acs{TotalEvalAccuracy}) \\
\end{tabular}
\label{table:manual-eval-stats}
\end{table}

\subsection{Production Adoption \& User Behavior}
After the manual evaluation, we launched \acs{AutoRepair} to automatically run on all changes with TAP test failures, \acs{TotalChanges} in total, and display suggested fixes across Google’s code repository, starting \acs{AutoRepairLaunchDate} continuing to date, with detailed statistics listed in Table \ref{table:agent-funnel}.

\begin{table}
\centering
\caption{Statistics about \acs{AutoRepair} executions and usage across Critique and Cider.}
\begin{tabular}{r|l}
\# changes with failing tests & \acs{TotalChanges} \\
\hline
\# changes dropped by pre-execution filters & \acs{TotalChangesDroppedPreExecution} \\
\# changes \acs{AutoRepair} attempted to fix & \acs{TotalChangesAttempted} (\acs{sym:funnel}\acs{TotalChangesAttemptedPct}) \\
\# changes \acs{AutoRepair} obtained a fix & \acs{TotalChangesAttemptedSuccessful} (\acs{sym:funnel}\acs{TotalChangesAttemptedSuccessfulPct}) \\
\# changes dropped by post-execution filters & \acs{TotalChangesDroppedPostExecution} \\
\# changes \acs{AutoRepair} posted a fix to & \acs{TotalChangesPostedTo} (\acs{sym:funnel}\acs{TotalChangesPostedToPct}) \\
\# changes developers previewed a fix & \acs{TotalChangesPreviewed} (\acs{sym:funnel}\acs{TotalChangesPreviewedPct})\\
\# changes developers applied a fix & \acs{TotalChangesApplied} (\acs{sym:funnel}\acs{TotalChangesAppliedPct})\\
\hline
\# changes with a fix previewed in Critique & \acs{TotalChangesPreviewedCritique} \\
\# changes with a fix applied in Critique & \acs{TotalChangesAppliedCritique} (\acs{CritiqueAppliedOverPreviewed}) \\
\hline
\# changes with a fix previewed in Cider & \acs{TotalChangesPreviewedCider} \\
\# changes with a fix applied in Cider & \acs{TotalChangesAppliedCider} (\acs{CiderAppliedOverPreviewed})
\end{tabular}
\label{table:agent-funnel}
\end{table}

\subsubsection{System Trigger}
Our system triggered on \acs{TotalChanges} changes with failing tests. \acs{TotalChangesDroppedPreExecution} were skipped by pre-execution filters.

A significant subset of the changes get skipped at this stage, as \acs{AutoRepair} targets human-authored, viable fix attempts. This points to future improvement opportunities to attempt fixing a wider set of code changes, including non-human authored, and larger changes.

\subsubsection{Fix Attempts}
Summarized in Table \ref{table:change-stats}, \acs{AutoRepair} attempted to fix \acs{TotalChangesAttempted} changes authored by \acs{TotalAuthors} distinct developers, there were a mean of \acs{ChangesMeanFiles} and median of \acs{ChangesMedianFiles} files per change, there were \acs{ChangesTotalFileExtensions} distinct file extensions across all changes, there were a mean of \acs{ChangesMeanTests} and median of \acs{ChangesMedianTests} failing tests per change.

\begin{table}
\centering
\caption{Statistics about code changes \acs{AutoRepair} attempted to fix.}
\begin{tabular}{r|l}
\# changes \acs{AutoRepair} attempted to fix & \acs{TotalChangesAttempted} \\
\# distinct authors of those changes & \acs{TotalAuthors} \\
\hline
mean \# files per change & \acs{ChangesMeanFiles} \\
median \# files per change & \acs{ChangesMedianFiles} \\
\# distinct file extensions (e.g. \texttt{.py}) across changes & \acs{ChangesTotalFileExtensions} \\
mean \# failing tests per change & \acs{ChangesMeanTests} \\
median \# failing tests per change & \acs{ChangesMedianTests} \\
\hline
mean \# tool calls per \acs{AutoRepair} execution & \acs{ChangesMeanToolCalls} \\
mean \# tokens used per \acs{AutoRepair} execution & \acs{ChangesMeanTokens} \\
total \# tokens used by \acs{AutoRepair} to date & \acs{ChangesTotalTokens} \\
\end{tabular}
\label{table:change-stats}
\end{table}

\acs{AutoRepair} made a mean of \acs{ChangesMeanToolCalls} tool calls, used a mean of \acs{ChangesMeanTokens} tokens in each of its executions, and a total of \acs{ChangesTotalTokens} tokens to date.

\begin{figure}
\centering
\includegraphics[scale=0.39]{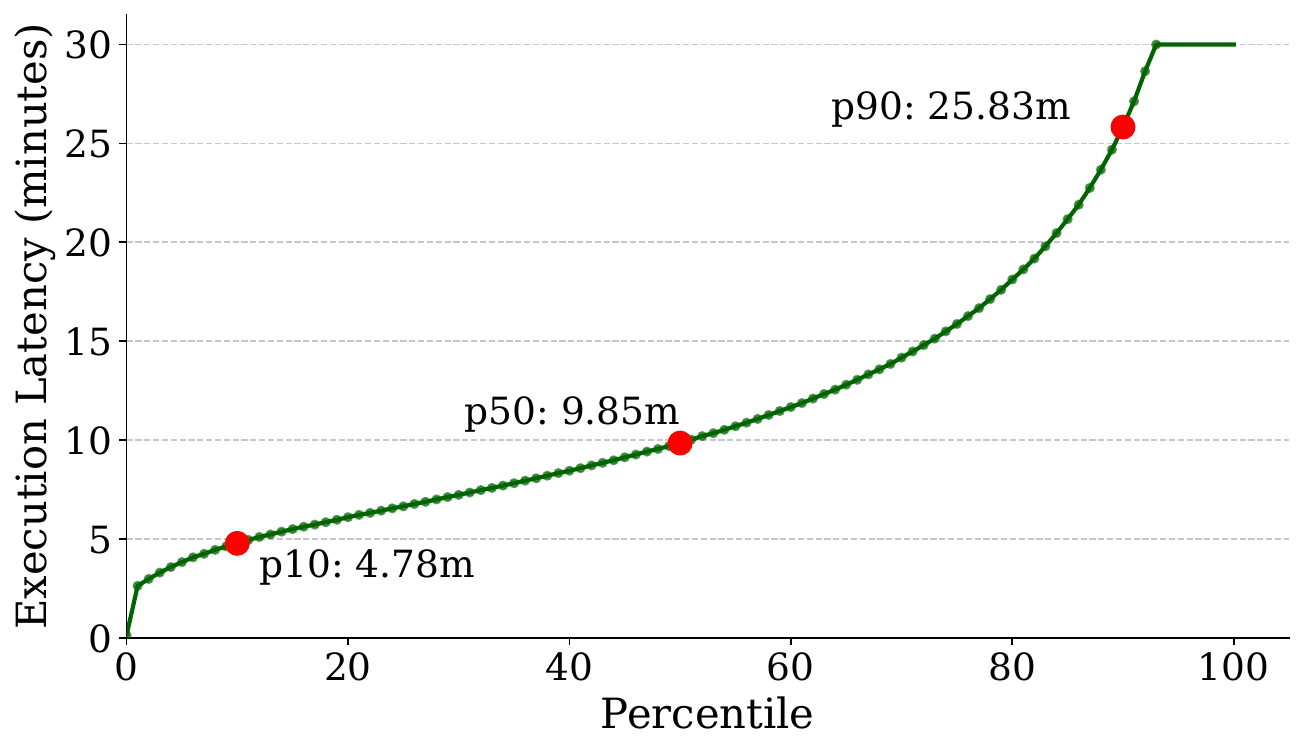}
\Description{\acs{AutoRepair} execution latency}
\caption{\acs{AutoRepair} execution latency.}
\label{fig:agent-execution-latency}
\end{figure}

Shown in Figure \ref{fig:agent-execution-latency}, from TAP failure notification to obtaining a successful fix, it took \acs{AutoRepair} \acs{AutoRepairLatencyP50} in the median (p50), and \acs{AutoRepairLatencyP90} in the 90th percentile (p90) wall clock time, within the median fix attempt latency of \acs{FixAttemptLatencyP50} discussed in Section \ref{sec:prior-sota}.

\acs{AutoRepair} successfully generated a suggested fix for \acs{TotalChangesAttemptedSuccessful} changes. Unsuccessful attempts are due to the limits placed on the agent execution to avoid running for too long, as users may get notified of the failing tests and attempt to fix them manually themselves.

Based on our internal experiments, we predict that a significant subset of the failed attempts at this stage may successfully produce fixes with improved language models released after our study.

\subsubsection{Successful but Dropped Fix Suggestions}
\acs{TotalChangesDroppedPostExecution} successful fixes were dropped by post-execution filters, as the developers modified or submitted their changes already, and the agent's fixes were obsolete.

Some of the fixes may still be relevant even though the developer modified the change, as they may have modified unrelated files. However, such fixes may confuse developers, and potentially cause conflicts with the developer's changes. As a result, they are dropped entirely by post-execution filters.

Even though \acs{AutoRepair} is optimized for low latency, a non-negligible share of successful fixes are still dropped by the post-execution filters, pointing to a need for further latency improvements to \textit{catch} developers in their change modification flows.

\subsubsection{Suggested Fixes}
After post-execution filters, \acs{TotalChangesPostedTo} fixes were posted to Critique. Across Critique and Cider, \acs{TotalChangesPreviewed} fixes were previewed and \acs{TotalChangesApplied} were accepted by developers.

\begin{figure}[tbp]
    \centering

    \begin{minipage}{\linewidth}
        \centering
        \includegraphics[width=\linewidth]{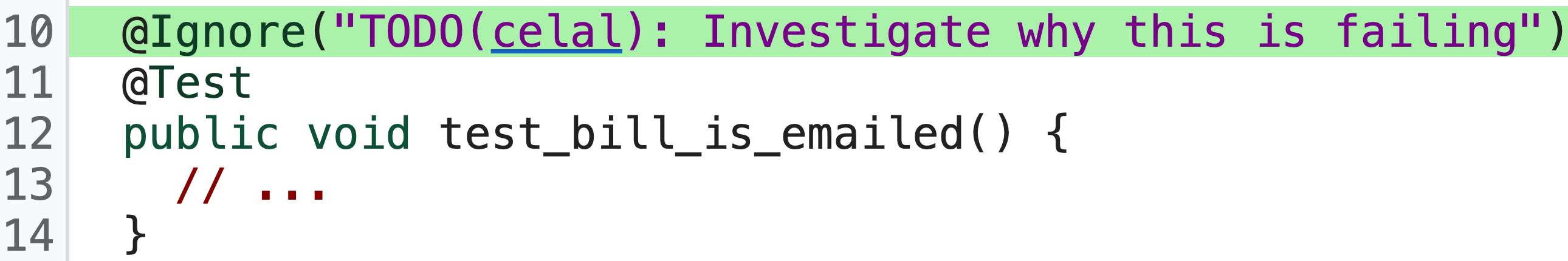}
        \subcaption{\texttt{@Ignore} added to test}\label{fig:bad-fix-examples-ignore}
    \end{minipage}
    
    \begin{minipage}{\linewidth}
        \centering
        \includegraphics[width=\linewidth]{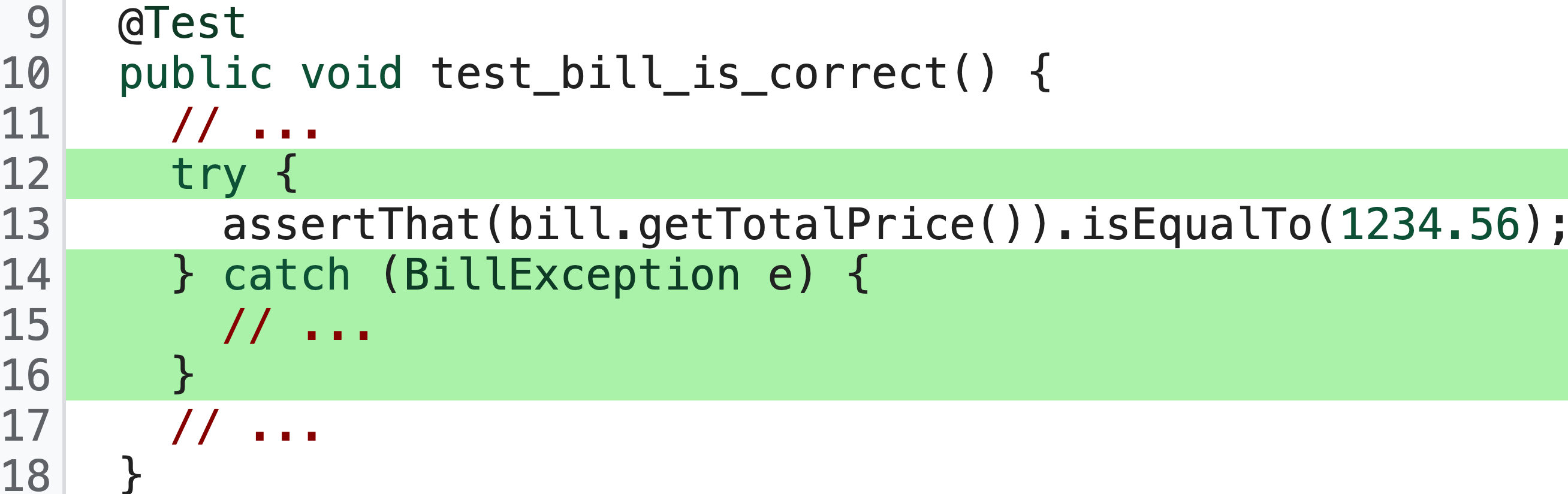}
        \subcaption{Assertion swallowed by try/catch block}\label{fig:bad-fix-examples-try-catch}
    \end{minipage}

    \begin{minipage}{\linewidth}
        \centering
        \includegraphics[width=\linewidth]{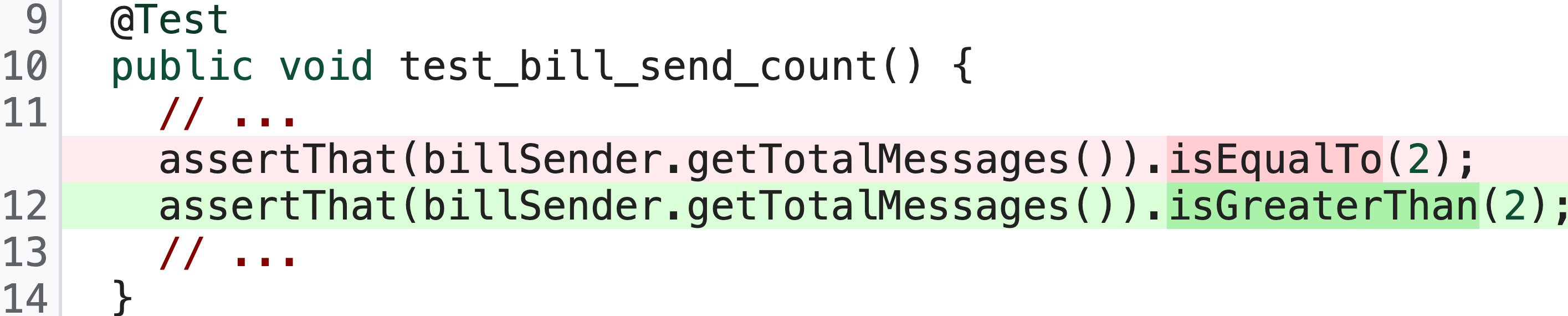}
        \subcaption{Fix is close but not exact, should assert "\texttt{== 3}", not "\texttt{> 2}"}\label{fig:bad-fix-examples-close-not-exact}
    \end{minipage}

    \Description{Examples of suggested fixes not applied by users.}
    \caption{Examples of suggested fixes not applied by users.}
    \label{fig:bad-fixes}
\end{figure}

Based on our manual analysis of a subset of previewed but not applied fixes of failing tests that have later been manually fixed by developers, we observe that there are several reasons why developers do not apply fixes, with examples shown in Figure \ref{fig:bad-fixes}.

First, even though the fix helps the tests pass, sometimes it is irrelevant or wrong, as shown in Figure \ref{fig:bad-fix-examples-ignore} and \ref{fig:bad-fix-examples-try-catch}. Such fixes should not be suggested to developers, as they undermine trust in the long run, and they should instead be verified with an additional step in our agent architecture, left as future work based on our study.

Second, there are cases where the fix is correct, but not exactly what the developer wants, as shown in Figure \ref{fig:bad-fix-examples-close-not-exact}. In these cases, developers most likely understand \textit{where} to fix the test as they preview the suggested fix, but then proceed to fix it themselves.

Then, there are cases where \acs{AutoRepair} suggested reverting the modifications in the change, essentially undoing developer's modification intent. Such suggestions should not be suggested to developers and have also been noted as future improvements to be addressed in the agent.

Finally, there are cases when the fix was correct, but developers still typed it manually themselves. We predict that those developers did not realize they could apply the fix in Critique, and are using an editor other than Cider, where the link posted in our findings is not applicable, or various other reasons, some of which are discussed in Section \ref{sec:user-interviews}.

\subsubsection{Developer Preferences on Applying Fixes}
In Critique, \acs{TotalChangesPreviewedCritique} fixes were previewed and \acs{TotalChangesAppliedCritique} fixes were applied, while in Cider, \acs{TotalChangesPreviewedCider} fixes were previewed and \acs{TotalChangesAppliedCider} applied.

Developers previewed some suggested fixes both in Critique and in Cider. Based on user interviews discussed in Section \ref{sec:user-interviews}, developers sometimes preview a fix in Critique, but then move to Cider to apply it to make further edits on it afterwards.

Additionally, \acs{AutoRepair} may suggest multiple fixes for a single change, as developers work through their change to satisfaction, and tests may fail on several versions of the change. This is the reason that the sum of the number of changes with a fix applied in Critique and Cider is greater than the number of changes developers applied a fix in Table \ref{table:agent-funnel}.

The $\frac{applied}{previewed}$ rate was \acs{CritiqueAppliedOverPreviewed} in Critique, while it was \acs{CiderAppliedOverPreviewed} in Cider. Developers seem to prefer viewing and applying fixes in Cider, compared to Critique. We propose that this is due to several factors, including developers' tendency to make further modifications on the suggested fixes, and developers being used to viewing diffs in the IDE, as supported by user interviews discussed in the next section.

\subsection{User Interviews}
\label{sec:user-interviews}
We conducted 45 minute in-person interviews with \acs{TotalInterviewedDevelopers} developers from \acs{TotalInterviewedDevelopers} different groups that have been working at Google for at least 6 months, that previewed and/or applied suggested fixes either in Critique or in Cider, with excerpts discussed below. In this section, we summarize key learnings based on the participants’ responses.

\subsubsection{Perceptions on \acs{AutoRepair}}
First, we asked participants about their perception on automatically receiving fix suggestions from \acs{AutoRepair} without triggering it themselves. Participants \acs{DeveloperF} and \acs{DeveloperG} reported enjoying the integration of \acs{AutoRepair} into existing development workflows they are accustomed to.

Then, we asked participants their prediction on how \acs{AutoRepair} works, and they all reported that it is most likely an AI-based product, as the fix suggestions it provides are not likely to come from a non-AI tool, and its name suggests it uses AI.

As a follow up, we asked them if this creates bias in their expectations of \acs{AutoRepair}'s suggested fixes. As \acs{DeveloperA} summarized it, participants reported that it does not matter where the suggestion comes from as long as it is relevant and accurate.

\begin{testimonialbox}
{[\acs{DeveloperG}]}: [...] I go to some other tasks in the mean time and then this new thing I've never seen before called \acs{AutoRepair} just fixes the test for me!\\[0.5em]

{[\acs{DeveloperF}]}: This [workflow] is really great. It’s in-line with a lot of Critique workflows, so it’s following the same pattern to not have to think what type of review it is because the approach is always the same.\\[0.5em]

{[\acs{DeveloperA}]}: I don't care what, who, or where the suggestion comes from, just that it is relevant and accurate.
\end{testimonialbox}

\subsubsection{Root Cause Summaries in Findings}
About the root cause summaries \acs{AutoRepair} findings provide, participants \acs{DeveloperD}, \acs{DeveloperE} and \acs{DeveloperF} reported easily understanding the failure reason by reading the root cause analysis summary in the finding.

\begin{testimonialbox}
{[\acs{DeveloperD}]}: It explains the issue pretty clearly. Thanks for building this! Can definitely see myself using this going forward :)\\[0.5em]

{[\acs{DeveloperE}]}: It pulled out the conclusion and put it at the top of the message, [...] know relatively fast glance what the error was; if this didn't exist, it would have taken longer to dig through what the actual error was.\\[0.5em]

{[\acs{DeveloperF}]}: This information is helpful; usually have to go into test logs, but now it's putting all the information in front, saves time; wouldn't have to go into logs, filter by error, don't need to scan all of that, have summarized info; can skip step of finding the issue.
\end{testimonialbox}

\subsubsection{Previewed and Applied Fixes}
Several participants, \acs{DeveloperA}, \acs{DeveloperG}, \acs{DeveloperH}, reported finding \acs{AutoRepair} fixes very useful, especially when the fix is on repetitive and tedious changes such as using mocks, and on refactoring related changes.

\begin{testimonialbox}
{[\acs{DeveloperG}]}: Did you guys implement \acs{AutoRepair} on Critique? This thing is amazing! It fixed [change]. I updated some code, then the test failed cause it was using a bunch of mocks. I hate updating these types of tests.\\[0.5em]

{[\acs{DeveloperA}]}: The experience of \acs{AutoRepair} fixing complex tests (especially those involving many mocks) was one of my biggest wow moments for LLM agents.\\[0.5em]

{[\acs{DeveloperH}]}: Thanks for developing and maintaining a tool which I find very promising and useful for developer velocity! Even my very first experience makes me excited about offloading the grungy task of looking at multiple documentation and past solved bugs to find easy parameter changes to fix tests.
\end{testimonialbox}

\subsubsection{Previewed but Not Applied Fixes}
Based on telemetry, we observed that, in some cases, developers previewed fixes but did not apply them, and we asked them about the reasons behind these.

Participants \acs{DeveloperA}, \acs{DeveloperB} and \acs{DeveloperC} reported that \acs{AutoRepair} sometimes suggested irrelevant or wrong fixes.

\begin{testimonialbox}
{[\acs{DeveloperA}]}: The fix was inaccurate, I didn’t want to sort the list.\\[0.5em]

{[\acs{DeveloperB}]}: [...] it suggested to add something that wasn't needed, so it wasn't fixing anything.\\[0.5em]

{[\acs{DeveloperC}]}: It was inaccurate, I didn’t think the backslash should be removed.
\end{testimonialbox}

Participants \acs{DeveloperB} and \acs{DeveloperC} reported that \acs{AutoRepair} findings sometimes arrive too late, by which point they have initiated their own manual fix. If the finding had come on time, they would have found it useful.

\begin{testimonialbox}
{[\acs{DeveloperB}]}: To be honest, I was frustrated by the broken [...] all day. I've manually debugged the problem, and it was cumbersome to find the root cause across so many different layers. I can confirm that your agent's output is correct, and it would have saved me a lot of time if I had seen it before manually digging through the logs.\\[0.5em]

{[\acs{DeveloperC}]}: The [finding] did successfully identify the issue! Or at least, one of the issues. It definitely would have helped.
\end{testimonialbox}

Several developers also reported that they sometimes preview the suggested fix, but instead of applying it in Critique or Cider, they type the fix in the IDE themselves manually, as the proposed fix does not exactly conform to their coding style, and they prefer a slightly different edit, similar to shown in Figure \ref{fig:bad-fixes}.

Participant \acs{DeveloperF} reported that they applied the fix manually to save time and avoid issues that may stem from the Critique workflow conflicting with changes they wanted to make in their IDE.

\begin{testimonialbox}
{[\acs{DeveloperF}]}: I have a second monitor, or third monitor depending on where I am working [...], but my Cider is already open there. So instead of going to click Show Fix, apply the fix, [...], I just move my mouse over [...] and amend my code, in case I get into a [conflict] or something weird.
\end{testimonialbox}

Finally, participant \acs{DeveloperI} reported that even if they fix the failures themselves, having an agent independently provide feedback is useful, as the agent has its own context on the intent of the change and the resulting failure.

\begin{testimonialbox}
{[\acs{DeveloperI}]}: I like having a separate agent in Critique to also chime in, as its context is fresh.
\end{testimonialbox}

Even though there are clear instructions in its system prompt, \acs{AutoRepair} sometimes suggests reverting the change or commenting out the failing tests, as shown in Figure \ref{fig:bad-fixes}. Several developers reported getting frustrated about such fix suggestions, and some made humorous statements about \acs{AutoRepair} both during our interviews and in internal forums. As this is a critical user trust problem, we are currently working on addressing it by adding an additional step into \acs{AutoRepair} in Figure \ref{fig:auto-repair-design} to verify that the suggested fix is not removing tests or reverting the developer's change, before sending the suggestion to the developer.

\subsubsection{Usability Feedback}
During the interviews, several participants asked for a feature to have an estimate of the remaining time until \acs{AutoRepair} finishes, so that they can decide whether to wait for its fix suggestion or to fix the failures themselves.

\begin{testimonialbox}
{[\acs{DeveloperD}]}: [...] if there’s an indicator that it’s running, I could say, "okay, it’s trying to fix some things", I could maybe go do something else and come back to this.
\end{testimonialbox}

Although we could not add the feature immediately, as it is not supported in Critique, we observe that developers are adapting to working with AI agents to take decisions on their waiting behaviors \cite{maister1985psychology,watkins2011my}. We are currently working on addressing this by adding additional mechanisms to notify users that \acs{AutoRepair} is working on the failures on their change.

\section{Threats To Validity}

In this section, we discuss the threats to the validity of our work and case studies.
\\

\noindent \textbf{Manual evaluation}: Our manual evaluation was conducted by three experienced developers. However, these developers did not own the production or test code of the failing tests, and may have made mistakes in their reporting of the accuracy of \acs{AutoRepair} fixes.

Additionally, the failures used in this evaluation were randomly chosen across Google, and may not be representative of all failures.
\\

\noindent \textbf{Filters}: The pre-execution filters are implemented based on early feedback from a small group of \acs{AutoRepair} users. As a significant set of changes are skipped due to these filters, our evaluation results may not generalize to the changes with such characteristics.
\\

\noindent \textbf{Developer bias}: \acs{AutoRepair} suggests fixes for all eligible code changes across Google. However, developers have varying positive/negative judgments on the use of AI and AI assistance in their development journeys. As we do not control for such variables in our evaluation of the previews and applies of the fixes, this may have introduced a selection bias, and biased the accuracy and validity of our results.

Additionally, the interviews were conducted with code change authors that volunteered to give feedback. This constitutes a self-selected group, users who did not participate in our surveys and interviews might have different perspectives that could alter the qualitative insights gained from those sessions.
\\

\noindent \textbf{Developer infrastructure}: Our case study was conducted within the specific context of Google's development environment and infrastructure. Although all internal tools have corresponding products in the industry, our findings may not be generalizable to other organizational settings.
\\

\noindent \textbf{LLM}: Our study uses an internal version of Gemini 2.5 Pro. Although popular large LLMs have been shown to perform well on tasks such as coding, the generalizability of our results to other LLMs may be limited.

Additionally, \acs{AutoRepair} relies on a specific prompt template for Gemini. Small changes in the prompt wording, structure, or ordering of instructions could potentially lead to different or degraded LLM performance. The robustness of the prompt engineering to minor variations or future LLM updates is a potential threat to the generalizability of our findings.
\\

\noindent \textbf{Tests}: As they are implemented by developers, the quality and detail of tests across different groups and systems at Google may vary significantly, and this may have influenced the accuracy of \acs{AutoRepair} on different tests and the results of our case studies.
\section{Related Work}

Automated program repair (APR) \citep{le2019automated} has received significant attention in literature, starting with non-machine learning based approaches, then machine learning based approaches, later moving on to LLM based techniques, and most recently LLM agent based methods.
\\

\noindent \textbf{Non-machine learning based APR}: Several techniques approach APR as a search problem with manually described mutation rules and repair patterns \citep{le2011genprog, le2016history, liu2019tbar}, or semi-automatic transformation rules derived from human implemented patches \citep{kim2013automatic, bader2019getafix, bavishi2019phoenix}. Some approaches use symbolic constraints to extract fixes \citep{nguyen2013semfix,ke2015repairing,xuan2016nopol,mechtaev2016angelix}, while others use static analysis to identify and then repair bugs \citep{van2018static,liu2023program,jain2023staticfixer}, and some methods replace faulty code with similar but correct code from the same codebase \citep{yang2022transplantfix}.

Beyond functional repairs, several approaches target different kinds of issues, including syntax errors \citep{wang2018search,gupta2019deep,sakkas2022seq2parse}, performance issues \citep{yu2018pinpointing}, security vulnerabilities \citep{harer2018learning}, type errors \citep{chow2024pyty}, and build breakages \citep{tarlow2020learning}.
\\

\noindent \textbf{Machine learning based APR}: Early work uses machine learning techniques to rank and choose fixes from among candidates \citep{long2016automatic}. Recent work proposes using machine learning to generate fixes, including neural machine translation models to map faulty code to fixed code \citep{gupta2017deepfix,tufano2019learning,lutellier2020coconut,chen2019sequencer}, tree transformation prediction models \citep{li2020dlfix,zhu2021syntax}, neural models for more specific bug types \citep{vasic2019neural}, and approaches that train models for specific repairs  \citep{ye2022neural,ye2022selfapr}. SelfAPR \citep{ye2022selfapr} trains a model using training samples obtained by perturbing a previous version of the program being repaired, enforcing the neural model to capture project specific knowledge.
\\

\noindent \textbf{LLM-based APR}: Recently, instead of training task-specific models, use of general purpose LLMs has been more common.

AlphaRepair \citep{xia2022less} is a zero-shot technique that framed APR as an in-filling task, without requiring any additional training data.

FitRepair \citep{xia2023automated} improves on AlphaRepair by incorporating the plastic surgery hypothesis \citep{barr2014plastic}, suggesting that the code needed to fix a bug typically exists within the same codebase already.

Several techniques use a one-shot prompt technique, where a single prompt is sent to the model with problem descriptions \citep{jiang2023impact,xia2023automated} or error messages \citep{joshi2023repair}, and the model responds with the proposed code fix. More recent techniques use iterative approaches, where the LLM is prompted several times and the feedback obtained from the previous repair attempts are passed to the subsequent iterations \citep{xia2024automated,kang2023explainable,ye2024iter,hidvegi2024cigar}. ChatRepair \citep{xia2024automated} is a dialogue based technique that works through a conversation with the engineer as it incrementally collects more information, proposes potential patches and generates better patches over time to locate and repair the bug. ITER \citep{ye2024iter} improves partial patches until they become plausible
and correct.

Agentless \citep{xia2024agentless} employs generated tests and a pre-defined three phase repair pipeline comprising of localization, repair and validation, achieving state-of-the-art performance at a remarkably low cost, outperforming previous open-source techniques.
\\

\noindent \textbf{Agent-based APR}: Most recently, new agentic approaches have emerged, where LLMs autonomously plan and perform tasks necessary to repair the program making use of tools provided to them.

FixAgent \citep{lee2024unified} uses specialized LLM agents ("Tester" and "Debugger") through prompt chaining for end-to-end fault localization and repair. LLM4FL \citep{nakhla2024enhancing} uses two LLM agents to iteratively analyze code to rank suspicious methods.

SWE-Agent \citep{yang2024swe} implements a ReAct-style \citep{yao2023react} open-ended loop with access to a set of APIs as tools to provide terminal access.

RepairAgent \citep{bouzenia2024repairagent} builds upon prior work on augmenting LLMs with tools to generate and execute code \citep{gao2022pal} invoked via APIs \citep{schick2023toolformer}, \citep{patil2023gorilla} and presents a state machine constrained agent, disallowing certain tool use sequences, unlike SWE-Agent.

AutoCodeRover \citep{zhang2024autocoderover} is an agent leveraging program metadata including class and method definitions. SpecRover \citep{ruan2024specrover} builds upon AutoCodeRover to introduce a natural-language specification for expected behavior and incorporates an agent that reviews potential patches to guide the repair agent.

CodeR \citep{chen2024coder} proposes a different architecture for APR, with a "manager" agent that coordinates sub-tasks described as a task graph.

MarsCode Agent \citep{liu2024marscode} proposes a dynamic, iterative APR approach with a generate-and-validate pipeline using multiple sub-agents.

OpenHands \citep{wang2025openhands} (previously named OpenDevin) is an open-source agent platform inspired by the Devin software engineering agent \citep{cognition2024introducing}, and provides an agentic foundation to build solutions for software engineering and other domains.

Google's Passerine \citep{rondon2025evaluating} is a ReAct-style \citep{yao2023react} agent employing tools including a test-suite-based metric calculator for verification, is deployed to production use, and is reported to be three times more successful on fixing machine generated bugs. Passerine has recently been improved with techniques to filter out patches unlikely to be accepted by humans \citep{cambronero2025abstain}.

Meta's Engineering Agent \citep{maddila2025agentic} also implements a ReAct-style \citep{yao2023react} loop to automatically fix failing unit-tests using an iterative loop to provide feedback on potential patches using validation and verification tools. Extending this work, Meta's MetaMateCR \citep{maddila2025aiassistedfixescodereview} provides AI-assisted fixes for reviewer comments.

Our work is an APR technique combining LLM-based and agent-based approaches discussed above. However, these prior work are focused on post-submit APR, i.e. repairs are performed offline in existing code without stringent low latency requirements. To the best of our knowledge, our work is the first one that triggers APR automatically upon failures in the pre-submit developer workflow, with latency requirements critical to the developer experience, and that reports outcomes and learnings in industrial scale for the pre-submit workflow respectively.
\\

\noindent \textbf{Benchmarks for APR}: The APR community created several public and private benchmarks to assess the quality of APR solutions.

HumanEvalFix \citep{muennighoff2023octopack} covers translated benchmarks for three software engineering downstream tasks, including program repair.

SWE-Bench \citep{jimenez2024swebench} is a benchmark sourced from GitHub issues across 12 popular open source projects, containing 2294 Python bugs and repairs. It also offers a smaller version, SWE-Bench-Lite \citep{swe-bench-lite}, with 300 bug/fix pairs. SWE-Bench Verified \citep{jimenez2024swebench} is a human annotated subset of SWE-Bench deemed "solvable" and is divided into “easy” and a “hard” subsets. SWE-Bench has emerged as the standard evaluation benchmark for many APR techniques. 

DebugEval \citep{yang2024enhancing} defines several software engineering tasks, including code repair, to evaluate LLMs' capabilities across languages.

DebugBench \citep{majdoub2024debugging} is a large dataset of LeetCode \citep{leetcode} coding problems with artificially injected bugs to compare LLMs on APR.

Recently, to address the complexity of creating benchmarks and facilitate comprehensive evaluations of APR methods, SWE-smith \citep{yang2025swe} has been proposed as a pipeline to generate software engineering tasks at scale, to support the automatic synthesis of code changes that break existing tests, and to drive the development of more robust and scalable APR solutions.

Another well-known benchmark is SWE-Lancer \citep{miserendino2025swe}, which consists of over 1400 freelance software engineering tasks.

Most recently, Google's Passerine \citep{rondon2025evaluating} uses an internal benchmark similar to SWE-Bench, while Meta's Engineering Agent \citep{maddila2025agentic} uses an internal machine generated tasks benchmark.

Our work has an internal benchmark comprised of real-world program failures at Google, discussed in Section \ref{section:evaluation}. We are not able to publicly release our benchmark dataset due to the constraints discussed in Section \ref{sec:mandatory-data}.
\section{Conclusion}
In this paper, we presented \acs{AutoRepair}, an autonomous AI coding agent deployed at Google to automatically debug and repair test failures during the pre-submit outer-loop workflow. By integrating directly with Google’s internal tools, Critique and Cider, \acs{AutoRepair} provides suggested fixes before developers switch context from their change modification flow. 

To ensure high-quality suggestions with low latency, the system employs pre-execution and post-execution filters, as well as an iterative generate-and-validate loop powered by an LLM.

Our evaluation demonstrates the efficacy and practical viability of this approach at an industrial scale. In a manual evaluation by \acs{TotalEvalDevelopers} expert developers, \acs{AutoRepair} successfully suggested accurate fixes for \acs{TotalEvalAccuracy} of evaluated test failures. Following its production deployment on \acs{AutoRepairLaunchDate}, the system suggested \acs{TotalChangesPostedTo} fixes, with developers accepting and applying \acs{TotalChangesApplied} of the suggestions.

Qualitative feedback from user interviews provides insights into adoption and user behaviors. Developers reported time savings and reduced frustration when diagnosing and fixing complex test failures. Challenges also remain, such as optimizing execution latency to catch developers before they context-switch, and mitigating unhelpful fix suggestions like commenting out tests.

Our findings underscore the potential of integrating LLM-based autonomous agents into latency critical industrial developer workflows for APR.

\section{Future Work}
We observed several improvement opportunities during our case studies of \acs{AutoRepair}.

First, we plan to improve \acs{AutoRepair}'s test deletion and code revert behaviors to avoid developer frustration.

Second, we plan to improve \acs{AutoRepair}'s latency and success to provide more fixes using more recent language models.

Additionally, we plan to add a feature to report to developers when \acs{AutoRepair} is working on fixing the failing tests on their changes, so that they can wait for it to finish.

Finally, we plan to extend \acs{AutoRepair} to AI-authored changes as well, as there is a general shift in the industry towards more AI-generated code changes.

\section*{Data Availability Statement}
\label{sec:mandatory-data}
The data that support the findings of this study are proprietary to Google and cannot be made publicly available. Due to strict confidentiality policies, commercial sensitivity, and non-disclosure agreements, we are unable to provide a public reproduction package or dataset.

\begin{acks}

We thank Gaurav Johari for his work on surveying the APR domain at Google, Adrian Berding, Ahmed Omran, Allen Edwards, Assaf Raman, Bernhard Konrad, Chris Lewis, Daotong(Daniel) Dai, Gaurav Johari, James Wilson, Jeyavaishnavi Muralikumar, John Penix, Mariana Stariolo, Nic Cornejo, Peter Spragins, Rob Dryke, Rose Rodrigues, Sara Toth, Sruthi Santhosh Nair, Will Findley, Yang Liu, Yurun Shen for their work on \acs{AutoRepair}, and Elaine Thai, Laurie Pham, Maria Arguello, Ryan McGarry for their help on conducting user interviews on \acs{AutoRepair} usage.

Google Gemini \citep{googleGemini2025} was utilized to generate sections of this work, specifically several tables in Latex \citep{lamport1994latex} format, and several graphs inside Google Colab \citep{googleColab}.

\end{acks}

\balance
\bibliographystyle{ACM-Reference-Format}
\bibliography{0-paper}
\end{document}